\documentclass[preprint,3p,times, twocolumn]{elsarticle}

\usepackage[ruled]{algorithm2e}
\usepackage{algpseudocode}
\usepackage{amssymb}
\usepackage{xcolor}
\usepackage{tikz}

\newcommand*\circled[1]{\tikz[baseline=(char.base)]{
            \node[shape=circle,fill,inner sep=1pt] (char) {\footnotesize \textcolor{white}{#1}};}}

\begin{document}

\begin{frontmatter}



\title{Kernel Fusion in Atomistic Spin Dynamics Simulations on \\Nvidia GPUs using Tensor Core}

\author[neu,sf,slac]{Hongwei Chen}
\author[ru]{Shiyang Chen}
\author[sf,slac]{Joshua J. Turner}
\author[neu]{Adrian Feiguin}

\affiliation[neu]{organization={Department of Physics, Northeastern University},
            city={Boston},
            postcode={02115}, 
            state={MA},
            country={USA}}
\affiliation[ru]{organization={Department of Electrical and Computer Engineering, Rutgers University},
            city={Piscataway},
            postcode={08854}, 
            state={NJ},
            country={USA}}

\affiliation[sf]{organization={Stanford Institute for Materials and Energy Sciences, Stanford University},
            city={Stanford},
            postcode={94305}, 
            state={CA},
            country={USA}}
\affiliation[slac]{organization={Linac Coherent Light Source, SLAC National Accelerator Laboratory},
            city={Menlo Park},
            postcode={94720}, 
            state={CA},
            country={USA}}

\begin{abstract}
In atomistic spin dynamics simulations, the time cost of constructing the space- and time-displaced pair correlation function in real space increases quadratically as the number of spins $N$, leading to significant computational effort. The GEMM subroutine can be adopted to accelerate the calculation of the dynamical spin-spin correlation function, but the computational cost of simulating large spin systems ($>40000$ spins) on CPUs remains expensive. In this work, we perform the simulation on the graphics processing unit (GPU), a hardware solution widely used as an accelerator for scientific computing and deep learning. We show that GPUs can accelerate the simulation up to 25-fold compared to multi-core CPUs when using the GEMM subroutine on both. To hide memory latency, we fuse the element-wise operation into the GEMM kernel using $\mathtt{CUTLASS}$ that can improve the performance by 26\% $\sim$ 33\% compared to implementation based on $\mathtt{cuBLAS}$. Furthermore, we perform the on-the-fly calculation in the epilogue of the GEMM subroutine to avoid saving intermediate results on global memory, which makes the large-scale atomistic spin dynamics simulation feasible and affordable. 
\end{abstract}




\end{frontmatter}


\section{Introduction}

Atomistic spin dynamics simulations provide valuable information about the energy spectrum of magnetic materials in different phases, allowing one to identify instabilities and characterize the nature of the magnetic excitations. The study of this field has been guided by theory and numerical simulations using Monte Carlo (MC) and atomistic spin dynamics methods \cite{takahashi1990dynamics,chen1994spin,samarakoon2017comprehensive,samarakoon2018classical,zhang2019dynamical,mohanta2020signatures,saha2021spin} to understand time-dependent non-equilibrium properties of magnetic phases and defects. These techniques are able to simulate the behavior of thousands of spins with unprecedented accuracy and faithfully reproduce experimental observations. An important measurement that provides information about the spectrum is provided by the spin correlations and, in particular, the dynamic structure factor $S(\mathbf{q}, \omega)$ that can be readily compared to experimental probes such as inelastic neutron and inelastic x-ray scattering \cite{Ament-2011-RMP}, or their time-domain counterpart $S(\mathbf{q}, t)$, through methods such as neutron spin echo or x-ray photon correlation spectroscopy \cite{MarkSutton-2008,sinha-advmat-2014}.

Previous performance studies for spin dynamics simulation mainly focus on the parallel implementation of the Monte Carlo method \cite{kaupuvzs2010parallelization,weigel2011gpu,komura2016improved,liang2017gpu,hassani2018parallelization} or solving the Landau-Lifshitz-Gilbert (LLG) equations\cite{evans2014atomistic,etz2015atomistic,ma2016spilady,tranchida2018massively,muller2019spirit} on multi-core CPUs and GPU. These strategies have been included in some open-source packages for spin dynamics simulation such as UppASD\cite{skubic2008method}, Vampire\cite{evans2014atomistic}, Spirit\cite{muller2019spirit}, and Sunny\cite{sunny}. To understand the dynamical properties, one usually needs to evaluate the space- and time-displaced pair correlation function and the dynamical spin structure factor, which can be obtained by spatial and temporal Fourier transform. In the case where the only degree of freedom is spin, spatial fast Fourier transform (FFT) can be utilized to perform calculations in momentum space to reduce time complexity. However, there are situations where periodicity is lacking, such as in molecules, or systems that undergo phase segregation into domains. Here, the total excitation spectrum needs to be calculated by summing the correlation function over all sites\cite{hellsvik2019general}. Moreover, in calculations that require obtaining $S(\mathbf{r}, \omega)$ \cite{samarakoon2018classical}, the spin-glass susceptibility \cite{de2005spatial,young2006numerical} or the microscopic correlation function\cite{baity2019mpemba} for a spin glass, for instance, the correlation function needs to be constructed in real space. However, the computational cost for obtaining the space- and time-dependent correlation function becomes the main bottleneck (see discussion below), significantly limiting the system sizes one can simulate.  In our previous study\cite{chen2023high}, we explored how to efficiently implement the computation of $S(\mathbf{q}, t)$ from real space spin-spin correlations function on x86 CPUs and proposed that the most computationally expensive term can be cast as matrix multiplications, and then accelerated by the highly optimized general matrix multiply (GEMM) subroutine. Still, the quadratic increase in computing hours on CPUs significantly limits the system sizes one can simulate. 

In this paper, we demonstrate the use of GPUs, which can massively parallelize linear algebra calculations, accelerating and scaling up the atomistic spin dynamics simulations. We first implement a straightforward method to calculate $S(\mathbf{q}, t)$ using a customized CUDA kernel with some optimizations for memory access, which serve as a baseline for later comparison. Then, we cast the time-dependent spin-spin correlation function as matrix multiplications that can be easily accelerated by calling highly optimized GEMM subroutine from $\mathtt{cuBLAS}$. To further improve the performance, we fuse the inner product into the GEMM kernel by loading the pre-calculated matrix to hide memory latency. The kernel fusion by loading strategy brings the best performance, but it can not be applied on large spin systems ($>40000$ spins) due to the memory limit. To enable large-scale simulations, the calculation of the $Q$ matrix used for the spatial Fourier transform and the inner product were performed in the epilogue of the GEMM to avoid saving these super large matrices ($N\times N$) on global memory. We show that the on-the-fly calculation can make the simulation of large spin systems in real space feasible on GPUs while maintaining better performance compared to $\mathtt{cuBLAS}$ based implementation.

\section{Background}
Magnetic materials can be modeled by the following generalized Heisenberg Hamiltonian
\begin{small}
\begin{equation} \label{eq:hamiltonian}
\mathcal{H} = \sum_{<ij>}J_{ij} \mathbf{S}_i\cdot \mathbf{S}_j -\sum_{<ij>}\mathbf{D}_{ij}\cdot (\mathbf{S}_i \times \mathbf{S}_j)  - A \sum_i (S^z_i) ^2 - H_z\sum_i S^z_i
\end{equation} 
\end{small}
where $S_i$ is a spin vector in the unit sphere, $\mathbf{D}_{ij}$ is the Dzyaloshinskii–Moriya interaction vector, $A$ is the single ion anisotropy, and $H_z$ the perpendicular magnetic field. The coupling $J$ can be positive (antiferromagnetic), favoring anti-parallel arrangements of spins, or negative (ferromagnetic), favoring configurations in which spins point in the same direction.

\subsection{Classical Monte Carlo}
 The finite temperature physics of this problem can be studied through classical Monte Carlo (MC) simulations using the Metropolis algorithm, a widely employed algorithm that has been the workhorse of the field for decades\cite{binder1973monte}. The detailed procedure is well-documented in existing literature\cite{binder1993monte,murthy2001introduction,landau_binder_2014}, and we hereby provide a brief summary. Spins are represented as classical vectors on the unit sphere. Given a spin configuration, each spin is updated by proposing a new trial move  randomly chosen on the sphere; the new spin orientation will be accepted or rejected using the von Neumann rejection method according to the probability $\mathbb{P}(\Delta E)=e^{(-\beta \Delta E)}$, where $\Delta E$ is the energy difference between the new and old spin configurations, and $\beta=1/T$ is the inverse temperature. This procedure is guaranteed to obey detailed balance and samples an equilibrium finite temperature distribution consistent with the partition function of the model\cite{binder1993monte}. At low temperatures, the acceptance rate of random walks may diminish significantly. A common strategy to mitigate this problem is to randomly generate the new trial spin within a small cone spanned around the initial spin. In our implementation, we adaptively adjust the size of the cone to maintain an acceptance rate of at least 20\%.

\subsection{Landau–Lifshitz spin dynamics}
To investigate the time-dependent dynamical behavior of classical spin models, Monte Carlo simulations and atomistic spin dynamics techniques need to be combined\cite{nowak2000monte}. MC simulations are used to generate spin configurations corresponding to the equilibrium distribution at a specified temperature. The temporal evolution of classical spins is described by the Landau-Lifshitz (LL) equation\cite{ellis2015landau}
\begin{equation} \label{eq:ll}
    \frac{ d \mathbf{S_i}(t)}{dt} = \frac{\partial \mathcal{H}}{\partial \mathbf{S_i}(t)} \times \mathbf{S_i}(t),
\end{equation}
where $\mathcal{H}$ is the Hamiltonian of the magnetic material and $\mathbf{S_i}(t)$ are the spin vectors at time $t$. With the spin trajectories $\mathbf{S_i}(t)$, the dynamic correlation function can be computed as 
\begin{equation} \label{eq:sqt}
    S(\mathbf{q}, t) = \frac{1}{N}\sum_{\alpha} \sum_{\mathbf{r}, \mathbf{r'}}e^{-i\mathbf{q}\cdot (\mathbf{r} -\mathbf{r'})}  C^{\alpha}(\mathbf{r} -\mathbf{r'}, t), 
\end{equation}
which is the spatial Fourier transform of the 
time-dependent spin-spin correlation function $C^{\alpha}(\mathbf{r} - \mathbf{r'}, t)$, defined as 
\begin{equation} \label{eq:ssc}
    C^{\alpha}(\mathbf{r} - \mathbf{r'}, t) = [\langle S^\alpha_{\mathbf{r}}(t) S^\alpha_{\mathbf{r'}}(0) \rangle - \langle S^\alpha_{\mathbf{r}}(t)\rangle \langle S^\alpha_{\mathbf{r'}}(0)\rangle ]. 
\end{equation}
The $\langle \rangle$ symbol represents the statistical average over thermalized realizations of the Monte Carlo simulation, and $\alpha$ represents the $x$, $y$, and $z$ components. The dynamic spin structure factor $S(\mathbf{q}, \omega)$ can then be obtained by temporal Fourier transforming $S(\mathbf{q}, t)$ to the frequency domain. In this work, we perform the conversion by utilizing the fast Fourier transform (FFT) and the windowing function in $\mathtt{NumPy}$ to minimize the ringing oscillations along the $\omega$ axis in $S(\mathbf{q},\omega)$.

\subsection{GPU and Tensor Core}
The GPU memory hierarchy comprises multiple forms of memory of different sizes and speeds. As an example, the A100 GPU has 40GB of high bandwidth memory (HBM) up to 2.0TB/s and 192KB of on-chip SRAM as shared memory per each of 108 Streaming Multiprocessors (SM) with bandwidth estimated around 19TB/s~\cite{abdelkhalik2022demystifying}.  Besides, each SM has 65,536 registers, which have the lowest access latency compared to other memory in GPU. The on-chip SRAM is an order of magnitude faster than HBM but many orders of magnitude smaller in size. Similarly, the register file is the fastest storage but is highly demanded, so the available size is much smaller than SRAM. As computing has gotten faster relative to memory speed~\cite{a100}, operations are increasingly cramped in a bottleneck by memory accesses. Thus, taking full advantage of the GPU memory hierarchy becomes important for realizing high-performance computing.

In recent GPU architectures, an SM often contains both general CUDA cores and tensor cores. For example, one SM in A100 GPU contains 32 double floating-point (FP64) cores and 4 tensor cores. Tensor cores are designed to perform matrix operations that are used extensively in scientific computing and machine learning. The third generation of tensor cores in A100 supports matrix operations in IEEE-compliant FP64 precision, and many scientific computing applications rely on it. One tensor core can perform 16 Fused Multiply Add (FMA) operations every cycle, and each tensor core instruction computes the GEMM between an $8\times 4$ matrix and $4\times 8$ matrix in 4 cycles. A large matrix multiplication problem can be accomplished by a collection of small matrix multiplication with the tensor core. The A100 GPU with 108 SMs delivers a peak FP64 throughput of 19.5 TFLOPS, which is 2x that of the regular CUDA core.

\subsection{CUTLASS}
$\mathtt{CUTLASS}$ (CUDA Templates for Linear Algebra Subroutines) is an open-source linear algebra library developed by NVIDIA for implementing high-performance GEMM \cite{cutlass}. The library decomposes the structure of the GEMM computation into modular software components for loading data, computing predicate masks, and streaming data at all levels and scales within CUDA with hardware-specific optimizations. It provides a collection of CUDA C++ templates and abstractions for the primitives at each level of the GEMM hierarchy to tune tiling sizes, data types, and other algorithmic policies. Besides, one exciting feature of $\mathtt{CUTLASS}$ is the support of matrix multiplication that runs on tensor cores using WMMA API\cite{wmma}. Therefore, one can leverage $\mathtt{CUTLASS}$ to develop the GEMM kernel comparable to the closed-source library like $\mathtt{cuBLAS}$ or $\mathtt{ cuDNN}$. The flexibility simplifies the development of customized GEMM-based kernels to deliver the best performance for many problems and applications\cite{huang2020strassen,ootomo2022recovering,osama2023stream}, especially in deep learning \cite{guo2020accelerating,kosaian2021arithmetic,chen2021re,zhai2022bytetransformer}.

\section{Implementations}
In this section, we present a step-by-step description of the multiple optimization procedures we implement to speed up the spin dynamics simulations and reduce memory usage to enable large-scale simulations with limited GPU memory. For convenience, we consider the calculation of $S(\mathbf{q}, t)$ for a ferromagnetic Heisenberg spin model on the 2D square lattice with only nearest-neighbor interaction $\mathcal{H} = J \sum_{<ij>} \mathbf{S}_i\cdot \mathbf{S}_j - H_z\sum_i S^z_i$, using $J=-1$ and $H_z=0.5$. Note that this Hamiltonian and lattice are chosen to demonstrate our performance optimizations, even though in this particular case the calculation can be carried out in momentum space to reduce computational cost. To validate the effectiveness of our optimizations, we evaluate the performance on Nvidia A100(40GB) SXM GPUs, and the CUDA program is compiled with $\mathtt{nvcc\ 11.7}$.

\subsection{Baseline}

\begin{algorithm}[h]
\caption{CUDA kernel function for solving LL equations}\label{alg:LL_cuda}
\KwData{$M$ spin configurations $\mathbf{S}_{i,r}(t)$ ($1 \leq  i \leq M$) }
\KwResult{$d\mathbf{S}_{i,r}(t)/dt$}
\tcp{Get position r for current thread}
$r \gets blockIdx * blockDimx + threadIdx $ \\
If $r \geq N$ \Return \\
\tcp{Perform calculation for $M$ spin configurations}
\For{$i \gets 1$ \KwTo $M$  } 
{
    \tcp{Effective field for position r}
    $effective\_field \gets 0$ \\
    \For{$j \gets 1$ \KwTo num\_neighbor}
    {
        $effective\_field \gets local\_exchange\_field$ \\
        $effective\_field \gets local\_dmi\_field$ \\
    }

    $effective\_field \gets anisotropy $ \\
    $effective\_field \gets magnetic\_field $\\
    \tcp{Final result}
    $d\mathbf{S}_{i,r}(t)/dt \gets local\_field \times \mathbf{S}_{i,r}(t)$
}

\end{algorithm}

The calculation of the dynamic correlation function can be separated into two parts, which are solving the LL equations in Eq.(\ref{eq:ll}), and computation of $S(\mathbf{q}, t)$ in Eq.(\ref{eq:sqt}). With $M$ MC realizations for $N$ spins systems, the time complexity of solving the LL equations is $O(M\cdot N)$ while the computation of $S(\mathbf{q},t)$ has a time complexity of $O(M\cdot N^2)$. The dynamical correlation function $S(\mathbf{q}, t)$ is computationally more expensive due to the summation over $r$ and $r'$ in the dynamical spin-spin correlation function $C^{\alpha}(\mathbf{r} - \mathbf{r'}, t)$. 

\begin{figure*}[h]
    \centering
    \includegraphics[width=0.65\textwidth]{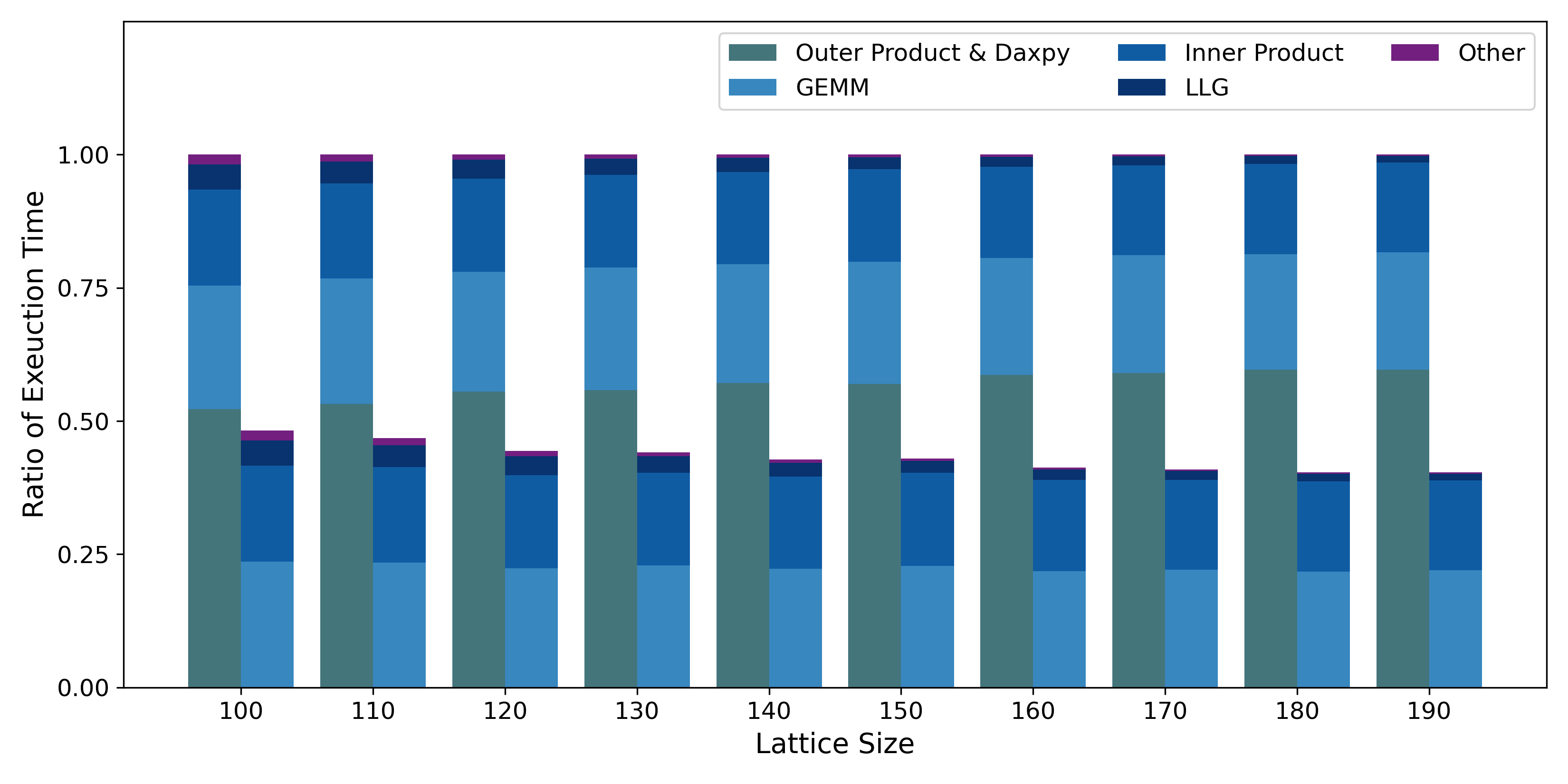}
    \caption{ We calculate the ratio of the execution time of each component of the first GEMM variant based on Eq.(\ref{eq:ssc}) (the left column) over the total execution time for different lattice sizes L($N=L^2$). The right column represents the second GEMM variant based on Eq.(\ref{eq:ssc_cov}). 50 MC realizations were used for the statistical average in all the implementations in this work. }
    \label{fig:ratio}
\end{figure*}

The time evolution in Eq.(\ref{eq:ll}) is computed using the fourth-order Runge-Kutta method, a general method for solving differential equations. While the evolution is not unitary, it is a cheap practical approach that serves its purpose for the current application. As the system evolves in time, the value of $ d \mathbf{S_i}(t) /dt$ is stored in a separate array so that there is no data dependence between different sites. For one MC realization, $\mathtt{CEIL(N/128)}$ thread blocks could be launched and each thread in the thread blocks performs the calculation of Eq.(\ref{eq:ll}) for one site. In our CUDA program, $M$ spin configurations were concentrated together and passed to the CUDA kernel function to avoid the overhead of launching the CUDA kernel multiple times. 
\begin{algorithm*}[h]
\caption{CUDA kernel function for calculation of $S(\mathbf{q},t)$ in baseline implementation}\label{alg:sqt_cuda}
\KwData{$M$ spin configurations $\mathbf{S}_{i,r}(t)$ at time $t$ ($1 \leq  i \leq M$) }
\KwData{$M$ spin configurations $\mathbf{S}_{i,r'}(0)$ at time zero ($1 \leq  i \leq M$) }
\KwData{Average of spin configurations $\langle S^\alpha_{\mathbf{r}}(t)\rangle$ at time $t$ }
\KwData{Average of spin configurations $\langle S^\alpha_{\mathbf{r'}}(0)\rangle$ at time zero }
\KwResult{$\sum_{r'} e^{-i\mathbf{q}\cdot (\mathbf{r} -\mathbf{r'})} C^{\alpha}(\mathbf{r} -\mathbf{r'}, t)$}
$num\_element \gets CEIL(\ N/blockDimx\ )$ \Comment{The number of elements for one thread} \\
$start\_pos \gets threadIdx*num\_element $\Comment{Get start position for current thread} \\
If $start\_pos \geq N$ \Return \\
Load $\mathbf{S}_{i, r}(t)$ ($1 \leq  i \leq M$) to shared memory\\
\_\_syncthreads() \\
\tcp{Divide the calculation of one row evenly in one thread block.}
\For{$r' \gets start\_pos\ \ \KwTo \ \ start\_pos+num\_element$  } 
{
     $C^{\alpha}(\mathbf{r} - \mathbf{r'}, t) \gets 0$ \Comment{Initialization}\\
    \For{$j \gets 1$ \KwTo $M$}
    {
        $C^{\alpha}(\mathbf{r} - \mathbf{r'}, t) \  \gets \  S^\alpha_{\mathbf{r}}(t) S^\alpha_{\mathbf{r'}}(0) $ \Comment{Calculate $\langle S^\alpha_{\mathbf{r}}(t) S^\alpha_{\mathbf{r'}}(0) \rangle$}
    }

    $C^{\alpha}(\mathbf{r} - \mathbf{r'}, t) \ \gets \langle S^\alpha_{\mathbf{r}}(t)\rangle \langle S^\alpha_{\mathbf{r'}}(0)\rangle $ \\
    Save $ S(\mathbf{q},t)_{r, r'} \gets e^{-i\mathbf{q}\cdot (\mathbf{r} -\mathbf{r'})}  C^{\alpha}(\mathbf{r} -\mathbf{r'}, t)$ to shared memory
}
\_\_syncthreads() \\
Reduce $ S(\mathbf{q},t)_{r, r'}$ for current thread block \\
\Return $\sum_{r'} e^{-i\mathbf{q}\cdot (\mathbf{r} -\mathbf{r'})} C^{\alpha}(\mathbf{r} -\mathbf{r'}, t)$
\end{algorithm*}

As for the calculation of $S(\mathbf{q},t)$, one straightforward method is to use a customized CUDA kernel function with a reduction kernel. $C^{\alpha}(\mathbf{r} - \mathbf{r'}, t)$ and $e^{-i\mathbf{q}\cdot (\mathbf{r} -\mathbf{r'})}$ can be recognized as 2D $N\times N$ matrices in memory, and $\sum_{\mathbf{r}, \mathbf{r'}}e^{-i\mathbf{q}\cdot (\mathbf{r} -\mathbf{r'})}  C^{\alpha}(\mathbf{r} -\mathbf{r'}, t)$ represents the element-wise product and then reduction (inner product). In the CUDA kernel function, each thread block is responsible for the calculation of elements in one row of the matrix $C^{\alpha}(\mathbf{r} - \mathbf{r'}, t)$, and $N$ thread blocks were launched. In each thread block, the computing tasks are evenly distributed between 256 threads, and the number of elements in $C^{\alpha}(\mathbf{r} - \mathbf{r'}, t)$ that was assigned to one thread is $\mathtt{CEIL(N/256)}$. For each element,  one $\mathtt{for}$ loop is necessary to compute the statistical average for the term $\langle S^\alpha_{\mathbf{r}}(t) S^\alpha_{\mathbf{r'}}(0) \rangle$. The computation of $e^{-i\mathbf{q}\cdot (\mathbf{r} -\mathbf{r'})} C^{\alpha}(\mathbf{r} -\mathbf{r'}, t)$ can then be carried out, and the results will be saved into shared memory to perform the reduction. Each thread block will return a scalar representing the inner product's result from one row. Then another reduction kernel is needed to compute the final result.

The above implementation represents a naive approach to evaluate $S(\mathbf{q},t)$, and we adopt several strategies to optimize the performance. Firstly, we compute three components $x$, $y$, $z$ of $C^{\alpha}(\mathbf{r} - \mathbf{r'}, t)$ simultaneously, taking advantage of the fact that the three components of spin vectors are stored continuously. Secondly, $\langle S^\alpha_{\mathbf{r'}}(0)\rangle$ were calculated and saved before the time evolution, as it remains constant throughout the simulation, eliminating the need to recalculate it at every step.  Also, a similar strategy can be applied on $\langle S^\alpha_{\mathbf{r}}(t)\rangle$ though it needs to be recalculated for each time $t$. We perform this operation before launching the customized CUDA kernel for $S(\mathbf{q},t)$. Thirdly, we optimize the memory access of $S^\alpha_{\mathbf{r'}}(0)$ by rearranging the order of data stored in global memory for $S^\alpha_{\mathbf{r'}}(0)$ to enable memory coalescing. Specifically, for each site $r$ from $1$ to $N$, the spin vectors $\mathbf{S}_{i,r}(t)(1 \leq  i \leq M)$ from $M$ Monte Carlo realizations were stored continuously instead of storing one whole spin configuration continuously. For $S^\alpha_{\mathbf{r}}(t)$, we first load the data from global memory to shared memory before the calculation in the CUDA kernel. Moreover, all the calculations were performed on registers since they are significantly faster.

\subsection{GEMM variant}

We tested the baseline implementation on Nvidia A100 for different sizes of 2D lattice using 50 MC realizations, finding that solving LLG only takes about 0.1\% $\sim$ 0.5\% of the total execution time, which means that the total run time is dominated by the computation of $S(\mathbf{q},t)$. Our previous work\cite{chen2023high} shows that solving the LL equation takes about 5\% of the overall simulation time in the baseline implementation on CPU, and $\langle S^\alpha_{\mathbf{r}}(t) S^\alpha_{\mathbf{r'}}(0) \rangle$ is the most expensive term which takes approximately 68\% $\sim$ 76\% of the total time. So, we adopt the same strategy to optimize the performance, which is to use the GEMM subroutine for $\langle S^\alpha_{\mathbf{r}}(t) S^\alpha_{\mathbf{r'}}(0) \rangle$ \cite{chen2023high}. The $\langle S^\alpha_{\mathbf{r}}(t) S^\alpha_{\mathbf{r'}}(0) \rangle$ term is not defined as matrix multiplication, but it can be cast as such, like the convolution operation\cite{chellapilla2006high} in deep learning.   

\begin{figure*}[ht] 
    \centering
    \includegraphics[width=0.75\textwidth]{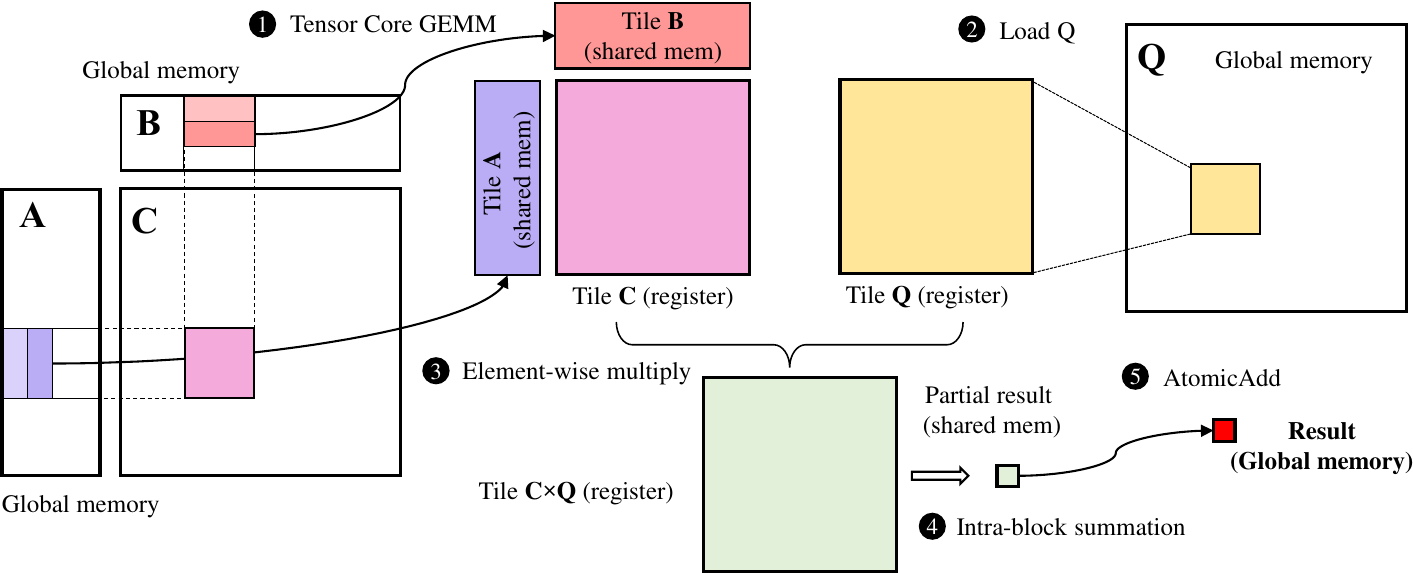}
    \caption{The computation diagram for thread blocks in the CUDA kernel that fuses GEMM and inner product.  Thread blocks first perform tensor core GEMM, and then load $Q$ matrix from global memory to perform the inner product.}
    \label{fig:kernel}
\end{figure*}

GEMM in $\mathtt{cuBLAS}$ employs a series of architecture-aware optimization strategies, such as tiling, warp-level tiling, register blocking, prefetching, mixed precision, and tensor core that improves the hardware utilization from a marginal $< 1$\% to a near-optimal efficacy ($>90\%$)\cite{huang2020strassen, zhang2017understanding, yan2020demystifying}. To leverage the highly optimized GEMM subroutine, the order of data in memory for spin configurations $S^\alpha_{\mathbf{r}}(t)$ at time $t$ needs to be rearranged. For each MC realization, we first store the $x$ component of spin vectors on every site, followed by the $y$ and $z$ components. The resulting rearranged spin configurations can be treated as a 2-dimensional matrix of size $N\times 3M$ stored in column-major. As for the spin configurations $S^\alpha_{\mathbf{r'}}(0)$ at time zero, the rearranged spin configurations in the baseline can be treated as a 2-dimensional matrix of size $3M\times N$ in column-major that meets the requirement. An additional matrix of size $N\times N$ is required to store the result of $\langle S^\alpha_{\mathbf{r}}(t) S^\alpha_{\mathbf{r'}}(0) \rangle$ computed by calling the GEMM subroutine. Another cost of using the GEMM subroutine is that the order of data for $S^\alpha_{\mathbf{r}}(t)$ needs to be rearranged for every step.

The computation of $S(\mathbf{q},t)$ can be decomposed into several basic linear algebra operations: GEMM for $\langle S^\alpha_{\mathbf{r}}(t) S^\alpha_{\mathbf{r'}}(0) \rangle$, outer product for $\langle S^\alpha_{\mathbf{r}}(t)\rangle \langle S^\alpha_{\mathbf{r'}}(0)\rangle$, Daxpy in Eq.(\ref{eq:ssc}), and inner product in Eq.(\ref{eq:sqt}). After the computation of $\langle S^\alpha_{\mathbf{r}}(t) S^\alpha_{\mathbf{r'}}(0) \rangle$ is finished, we perform the outer product and Daxpy in one customized kernel function to avoid saving the intermediate results. Similar to the baseline implementation, $N$ thread blocks were launched and each thread block is responsible for the calculation of one row. The inner product is calculated by calling $\mathtt{inner\_product}$ function in $\mathtt{Thrust}$. 

In Fig.(\ref{fig:ratio}), we calculate the ratio of the execution time of different components in the spin dynamics simulation over the total execution time based on Eq.(\ref{eq:ssc}) (the left column) on different lattice sizes L($N=L^2$). We can see that the most expensive component turns out to be the customized kernel for the outer product and Daxpy, which takes more than 50\% of the total time. This may be caused by the inefficient design of the customized kernel. In our CPU implementation\cite{chen2023high}, Daxpy and the inner product were fused into the GEMM kernel. However, there is an easy approach to hide the latency caused by the outer product and Daxpy, which is to rewrite the Eq.(\ref{eq:ssc}) in the following form, 
\begin{equation} \label{eq:ssc_cov}
    C^{\alpha}(\mathbf{r} - \mathbf{r'}, t) = \langle ( S^\alpha_{\mathbf{r}}(t) - \langle S^\alpha_{\mathbf{r}}(t)\rangle) (S^\alpha_{\mathbf{r'}}(0) -  \langle S^\alpha_{\mathbf{r'}}(0)\rangle )\rangle. 
\end{equation}
With ``updated'' matrices for $S^\alpha_{\mathbf{r}}(t) - \langle S^\alpha_{\mathbf{r}}(t)\rangle$ and $S^\alpha_{\mathbf{r'}}(0) -  \langle S^\alpha_{\mathbf{r'}}(0)\rangle $, the dynamical spin-spin correlation can be calculated by a single GEMM subroutine. $S^\alpha_{\mathbf{r'}}(0) -  \langle S^\alpha_{\mathbf{r'}}(0)\rangle $ only needs to be calculated once before the time evolution as it remains constant. The ``updated" spin configurations $S^\alpha_{\mathbf{r}}(t) - \langle S^\alpha_{\mathbf{r}}(t)\rangle$ at time $t$ can be calculated during the process of rearranging the order of data, and the time cost of this additional step is negligible to the end-to-end performance. The right column in Fig.(\ref{fig:ratio}) shows the ratio of the execution time of different components in the new implementation, where the total time from the first GEMM implementation (the left column) was used to calculate the ratio. We can see that the new implementation based on Eq.(\ref{eq:ssc_cov}) completely hides the computation time of the outer product and Daxpy without increasing the other steps' time cost. Eq.(\ref{eq:ssc}) and Eq.(\ref{eq:ssc_cov}) are mathematically equivalent, but the latter one is more computationally efficient. For this reason, in the following discussion, we use the second approach to represent the performance of the GEMM variant. 

\begin{figure*}[h]
    \centering
    \includegraphics[width=0.8\textwidth]{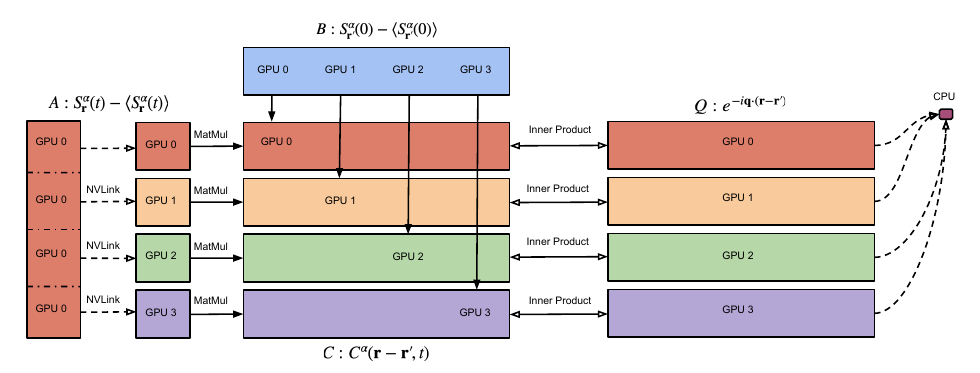}
    \caption{Example of dividing the computation of $S(\mathbf{q},t)$ between 4 GPUs on a single computing node. GPU 0 solves LL equations, and then the matrix $S^\alpha_{\mathbf{r}}(t) - \langle S^\alpha_{\mathbf{r}}(t)\rangle$ is evenly distributed on 4 GPUs. Each GPU executes the fused GEMM kernel (``Kernel Fusion: OTF\_SM''), and sends its result back to CPU. }
    \label{fig:multigpu_diagram}
\end{figure*}

\subsection{Kernel fusion}
From Fig.(\ref{fig:ratio}), we also observe that the inner product, which is a memory-bounded element-wise operation, also consumes non-negligible execution time. Since this operation is performed immediately after a computing-bounded operation, fusing the inner product into the GEMM kernel will hide the memory latency. In particular, at the end of the GEMM kernel, each thread block computes a tile of the resultant matrix in registers. Instead of storing it out to off-chip memory, a thread block loads the corresponding tile of the $e^{-i\mathbf{q}\cdot (\mathbf{r} -\mathbf{r'})}$ (Q matrix) to registers as well, in which we perform the element-wise addition for the partial results of the inner product. The partial results of each thread block are summed to derive the final result of the inner product. By fusing the kernel, we avoid the round trip of storing the GEMM result and loading it for the next kernel. Besides, the intermediate GEMM results do not require memory allocation, which reduces memory usage as well. 

To support kernel fusion, we leverage the open-sourced $\mathtt{CUTLASS}$ to implement the GEMM kernel since $\mathtt{cuBLAS}$ is closed-sourced. Fig.(\ref{fig:kernel}) illustrates the procedure of the fused kernel.  Step \circled{1} performs the tensor core GEMM. We partition the resultant matrix into tiles and assign each thread block one. The thread block collectively loads the required tiles from global memory to on-chip shared memory. We leverage the templates in CUTLASS to maximize bandwidth utilization by pre-computing the pointer offsets and coalescing memory transactions. We then iterate the tile in shared memory to load the sub-tiles to registers, from which the tensor core performs the matrix-multiply between sub-tiles, which also uses CUTLASS's modules. In step \circled{2}, we load the $Q$ matrix to the registers. Note that the register tiles in the example are shared by all threads in the thread block, in which each thread holds a part of it~\cite{ptx}. Each thread hereby loads and multiplies the elements in $Q$ correspondingly to its owning results as step \circled{3}. Specifically, the tile sizes of GEMM for each thread block and each warp are 128$\times$128 and 32$\times$64, respectively. We then sum the result of each thread hierarchically in step \circled{4}. We utilize the warp shuffle instructions and shared memory to exchange the thread registers for intra-block summation. Last, we use atomic operations to accumulate the thread block results to the global memory in step \circled{5}. This implementation is labeled as ``Kernel Fusion: Load" in the following comparison.   

\subsection{On-the-fly computation}
Although the complex $Q$ matrix is constant throughout the computation, the memory usage of this matrix will exceed the memory size of a single A100 GPU for $N>200$. To overcome this limitation, we propose the on-the-fly computation within our fused GEMM kernel to further reduce memory usage.  This approach requires recomputing the $Q$ matrix every time inside the kernel. Specifically, in step \circled{2} of Fig.(\ref{fig:kernel}), instead of loading the elements in the $Q$ matrix, threads compute  the following
\begin{equation} \label{eq:Q}
   Q(i, j) = e^{-i\mathbf{q}\cdot (\mathbf{r_i} -\mathbf{r_j})}.
\end{equation}
The time complexity of extra computation is $O(N^2)$, which is asymptotically smaller than the dominating GEMM complexity $O(M\cdot N^2)$.

\begin{figure*}[h]
    \centering
    \includegraphics[width=0.8\textwidth]{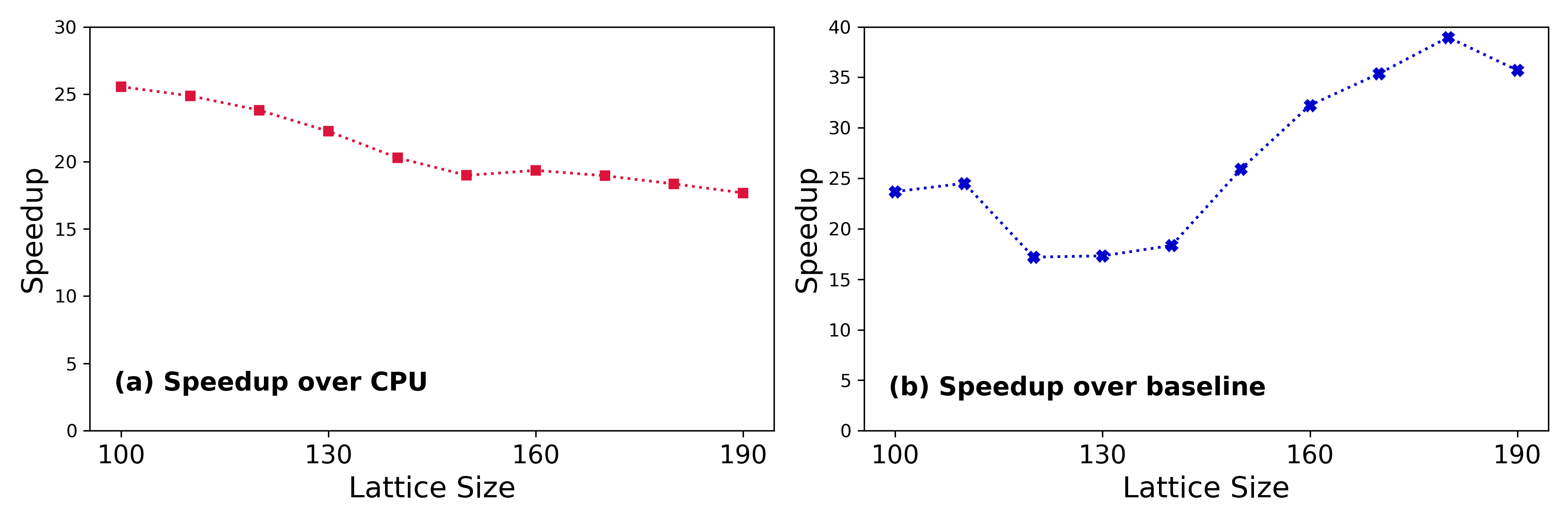}
    \caption{ Computational speedup of the GEMM implementation based on Eq.(\ref{eq:ssc_cov}) over CPU and baseline implementation. In the CPU implementation, calculations were performed on a 64-core AMD EPYC 7763 CPU using GEMM subroutine from Intel MKL.}
    \label{fig:speedup}
\end{figure*}

Unfortunately, the performance degrades because of the register spilling when computing the transcendental functions in Eq.(\ref{eq:Q})  (We label it as ``Kernel Fusion: OTF\_REG"). Our kernel heavily uses registers to store the frequently used matrix tiles, which can increase the computation efficiency in GEMM. Consequently, when calling transcendental functions, the compiler offloads the tiles in stack frames to off-chip memory. The storing and recovering of the stack frames leads to significant overhead due to memory latency. To tackle this issue, we store the Tile $C$ to shared memory after step \circled{1} to reduce the register usage footprint (We label it as ``Kernel Fusion: OTF\_SM''). Of note, we can reuse the shared memory for Tile $A$ and $B$ so that the total required memory does not increase too much. The elements in $Q$ are computed in registers and multiplied with values in shared memory. Therefore, we avoid materializing the full Tile $C\times Q$ in registers. Besides, we also use smaller tile sizes in GEMM to reduce resource usage. Specifically, both OTF\_REG and OTF\_SM use tiles size of 128$\times$64 for thread block and 32$\times$32 for warp.

 \subsection{Multi-GPU implementation}
To support the simulation for even larger spin systems, there is a need to scale the scheme to multiple GPUs. Fig.(\ref{fig:multigpu_diagram}) depicts the example of our multi-GPU implementation on a single node with 4 GPUs. Since the computation time of solving LL equations only takes less than 1\% of the total time, we consider a simple realization for multi-GPU, which only offloads the computation of $S(\mathbf{q},t)$ to the other three GPUs. First, GPU 0 computes matrix $A$ locally and splits the rows into 4 parts denoted as $A_1$ to $A_4$. The sub-matrix $A_i$ is sent to $i$-th GPU via NVLink when $i$ is not 0. Secondly, $i$-th GPU uses our fused kernel to compute the GEMM and inner product between $A_i$ and $B$. Because matrix $B$ is constant, each GPU possesses a replica before time evolution to reduce communication.  Each GPU stores its result in GPU memory and aggregates it back to GPU 0. Generally, when using $P$ GPUs, for each step of time evolution, GPU 0 sends data in the size of $3MN(P-1)/P$, and each GPU receives a sub-matrix with size $3MN/P$.  Therefore, our design has good scalability because the communication volume does not increase as the number of GPUs increases. 

\section{Results}

\subsection{Single GPU performance}

We first compare the performance of the GEMM variant with the baseline implementation on GPU and also the GEMM implementation on CPU for different 2D square lattices, where $L$ ranges from 100 to 190. For the CPU performance, the evaluation was performed on a 64-core AMD EPYC 7763 CPU using GEMM subroutine from Intel MKL. From Fig.\ref{fig:speedup}(a), we see that spin dynamics simulation on GPU is up to 25 times faster than CPU when running the same algorithm. Fig.\ref{fig:speedup}(b) shows that using the GEMM subroutine on GPU can speed up the calculation up to 39-fold compared to baseline.

\begin{figure*}[h]
    \centering
    \includegraphics[width=0.7\textwidth]{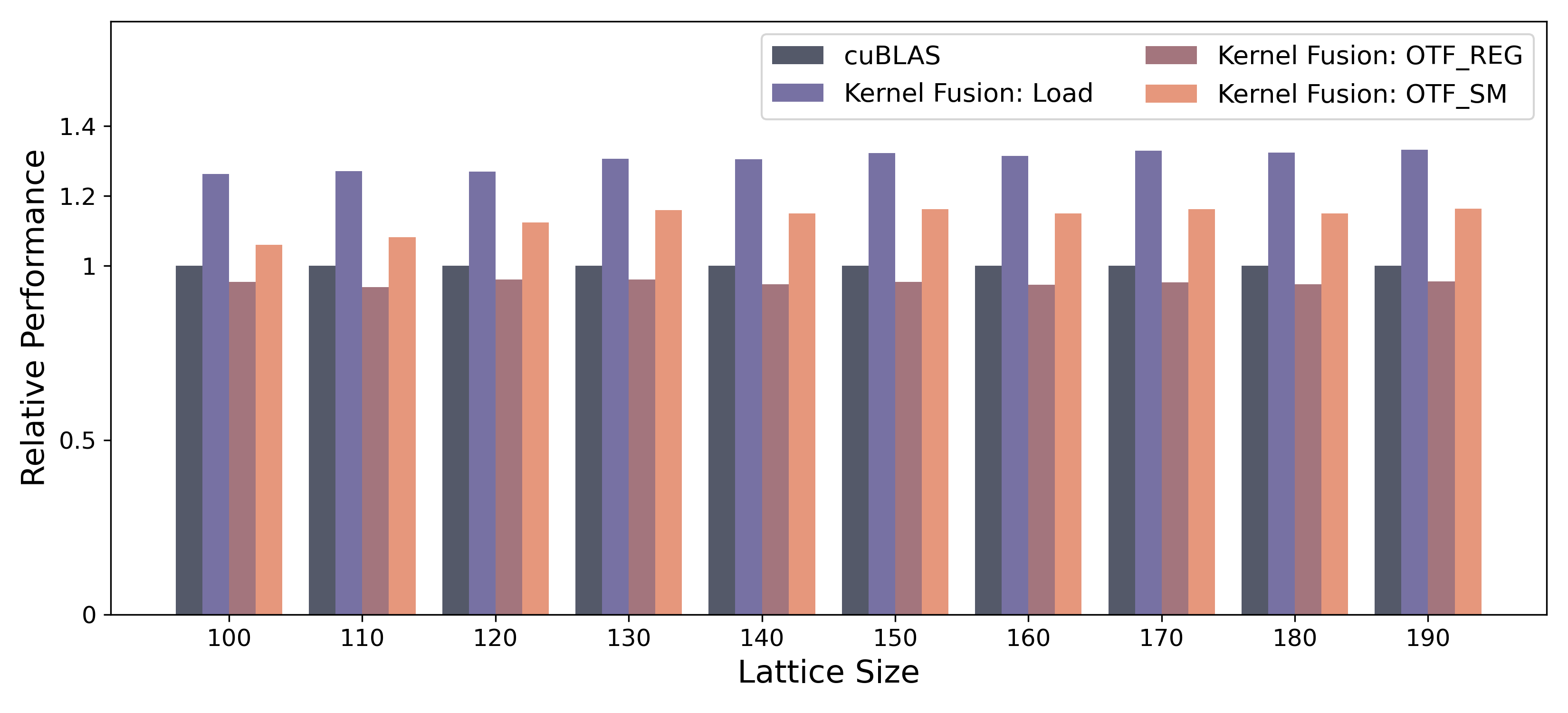}
    \caption{Relative performance improvement of 3 kernel fusion variants, ``Kernel Fusion: Load", ``Kernel Fusion: OTF\_REG" and  ``Kernel Fusion: OTF\_SM'',  over the GEMM variant ($\mathtt{cuBLAS}$) on different lattice sizes L($N=L^2$).  }
    \label{fig:perf}
\end{figure*}

Casting the time-dependent spin-spin correlation $C^{\alpha}(\mathbf{r} - \mathbf{r'}, t)$ as matrix multiplication greatly simplifies the optimization procedures, and the calculation of $S(\mathbf{q},t)$ can be treated as two basic linear algebra operations, which are GEMM and inner product. We fuse the inner product into the GEMM kernel using $\mathtt{CUTLASS}$, and it improves the performance by 26\% $\sim$ 33\% compared to $\mathtt{cuBLAS}$ and $\mathtt{Thrust}$ based implementation as shown in Fig.(\ref{fig:perf}). Note that the maximum lattice size one can simulate using $\mathtt{cuBLAS}$ and $\mathtt{Thrust}$ on an A100(40GB) GPU is smaller than $200\times 200$ because of the limited GPU memory. Fusing the inner product into the GEMM kernel avoids allocating global memory for the intermediate results from GEMM, but the $Q$ matrix ($N\times N$) still needs to be stored in global memory. 

To address the issue of memory limitation, we perform on-the-fly calculations for the spatial Fourier transform in the GEMM kernel. It avoids global memory allocation for the $Q$ matrix by calculating it in the GEMM kernel, but the calculation needs to be repeated when the customized GEMM kernel is called, which increases the workload and reduces the end-to-end performance. In Fig.(\ref{fig:perf}), we show that by saving the intermediate results of the $Q$ matrix on shared memory, our GEMM kernel with on-the-fly calculation (Kernel
Fusion: OTF\_SM) still outperforms the $\mathtt{cuBLAS}$ and $\mathtt{Thrust}$ based implementation by 5\% $\sim$ 16\%.  With the cost of losing 10\% $\sim$ 20\% of the end-to-end performance when compared to kernel fusion by loading variant, our strategy breaks through the bottleneck of simulations for large spin systems on GPU.

\begin{figure*}[h]
    \centering
    \includegraphics[width=0.7\textwidth]{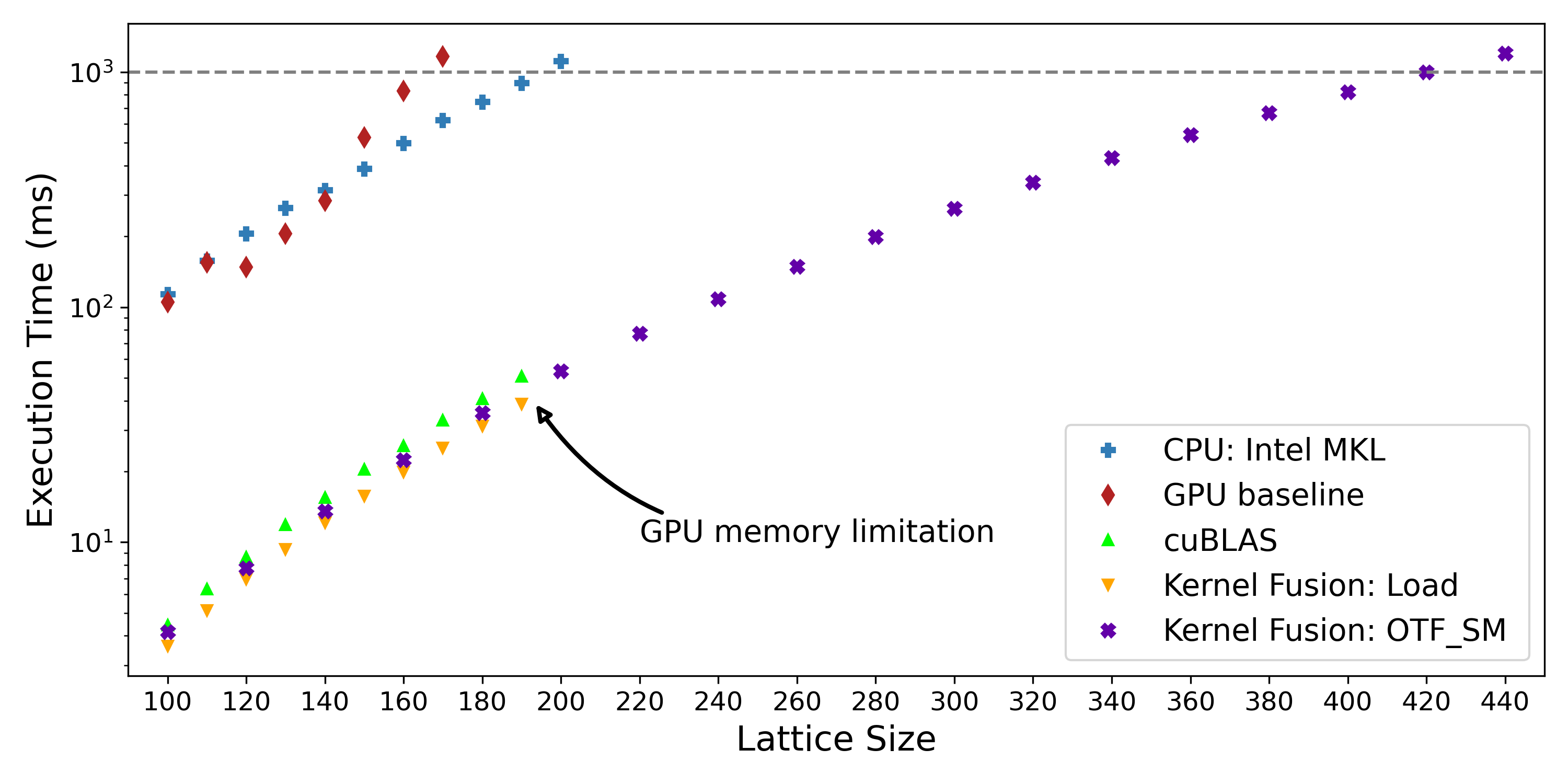}
    \caption{ Execution time of several implementations in this work for different lattice sizes $L$ ($N=L^2$). GPU memory limits the maximum lattice size that the GEMM variant ($\mathtt{cuBLAS}$) and kernel fusion by loading variant can calculate.}
    \label{fig:time}
\end{figure*}

\subsection{Multi-GPU performance}

\begin{figure*}[ht] 
    \centering
    \includegraphics[width=0.7\textwidth]{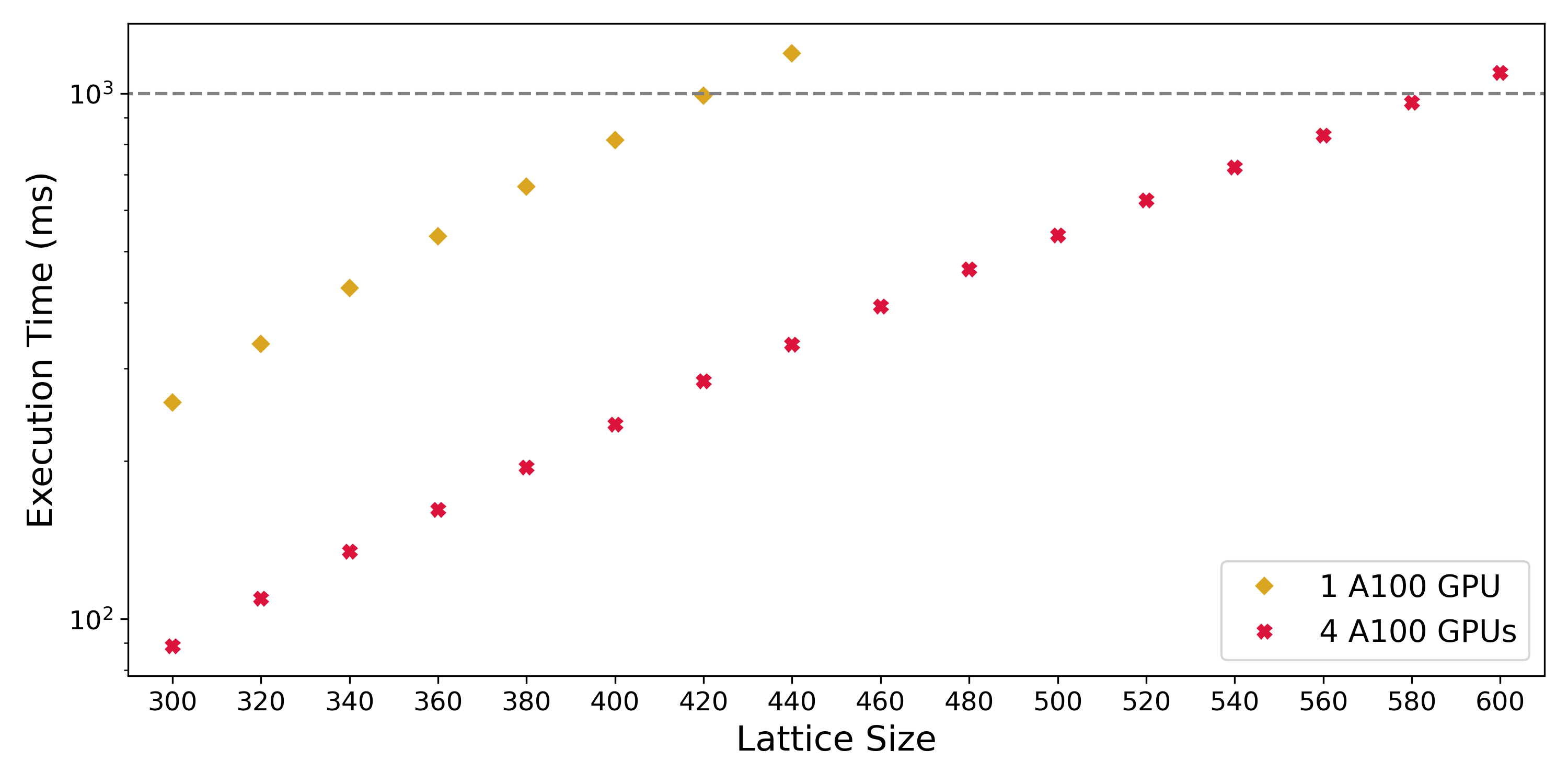}
    \caption{ Execution time for running on-the-fly calculation in the GEMM kernel (Kernel Fusion: OTF\_SM) using 1 and 4 GPUs for different lattice sizes $L$ ($N=L^2$).}
    \label{fig:multigpu_time}
\end{figure*}

The performance evaluation of the multi-GPU implementation was conducted on a single GPU computing node equipped with 4 A100 (40GB) SXM GPUs. Each pair of GPUs were interconnected by 12 third-generation NVLinks, with each link offering a bi-directional bandwidth of 50 Gb/s\cite{nvlink}. The time taken for the calculation of $S(\mathbf{q},t)$ for varying lattice sizes are shown in Fig.(\ref{fig:multigpu_time}). Under a 1000 ms constraint, 4 GPUs managed to finish the calculation of one step time evolution for a $580\times 580$ square lattice, while a single GPU was limited to a $420 \times 420$ square lattice. The speedup of using 4 GPUs, compared with one GPU on lattice size between 300 and 440, ranges from 2.9 to 3.6 fold. 

\subsection{Application}

\begin{figure*}[ht]
    \centering
    \includegraphics[width=0.8\textwidth]{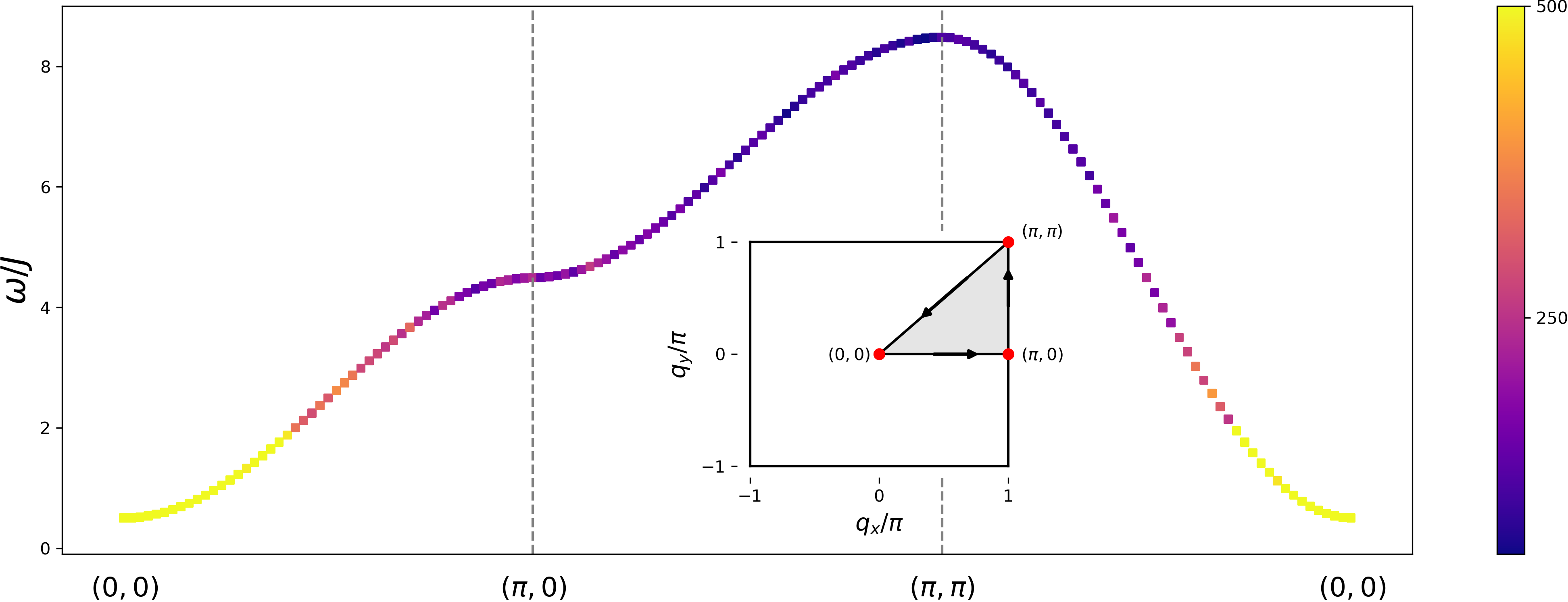}
    \caption{ Dynamic spin structure factor $S(q, \omega)$ calculated in the ferromagnetic phase  ($T=0.01J$) on a square lattice of size 100 × 100 with periodic boundary conditions, along the momentum path $(0, 0) \rightarrow (\pi, 0) \rightarrow (\pi, \pi) \rightarrow (0, 0)$. The inset shows the high symmetry path in the first Brillouin zone for the square lattice.}
    \label{fig:sqw}
\end{figure*}

In addition to performance optimization, we also apply our implementation to calculate the dynamic spin structure factor $S(q, \omega)$ for the ferromagnetic Heisenberg model. Using classical Monte Carlo and simulated annealing, we obtain 50 spin configurations of the 2-dimensional square lattice of size $100\times 100$ with periodic boundary conditions. We also use simulated annealing starting from a high temperature $T_0=10J$, and then incrementally decreasing it to the target temperature $0.01J$ using $T_{i+1} = 0.995T_i$. At every temperature, $10^5$ Monte Carlo sweeps ($10^9$ spin flips) are performed, and additional $5\times 10^5$ Monte Carlo sweeps were carried out at the target temperature to achieve equilibrium.  The dynamical spin structure factor $S(\mathbf{q}, \omega)$ was calculated along the high symmetry path of the first Brillouin zone using Eq.(\ref{eq:ll}) and Eq.(\ref{eq:sqt}), as shown in Fig.(\ref{fig:sqw}). 

\section{Conclusion}

In this work, we explore several implementations of atomistic spin dynamics simulations on GPUs to achieve high performance and reduce memory usage for large-scale simulations. Specifically, the calculation of the spin-spin correlation function is the most computationally expensive part of the simulation, and the performance of evaluating this term can be  easily optimized by casting it as matrix multiplication. Compared to the baseline, using GEMM subroutine from $\mathtt{cuBLAS}$ can speed up the calculation up to 39-fold. We also compare the performance with a GEMM implementation on CPU, finding that using an A100 GPU for such simulation is up to 25 times faster than a 64-core AMD EPYC 7763 CPU. To further improve the performance, we fuse the memory-bounded operation, the inner product, into the calculation of spin-spin correlation, which has been cast as matrix multiplication, to hide the memory latency. The kernel fusion strategy improves the performance by 26\% $\sim$ 33\% compared to $\mathtt{cuBLAS}$ and $\mathtt{Thrust}$ based implementation. To perform large-scale simulations, storing the $Q$ matrix will become a bottleneck because of the limited GPU memory. To address this issue, we perform on-the-fly calculations for the $Q$ matrix in the epilogue of the GEMM kernel to hide memory usage. By storing the intermediate results of the $Q$ matrix on shared memory, our fused GEMM kernel based on $\mathtt{CUTLASS}$ still outperforms the $\mathtt{cuBLAS}$ and $\mathtt{Thrust}$ based implementation by 5\% to 16\%.  We also demonstrated that our fused GEMM kernel can be adapted to perform such simulations on multi-GPUs. We highlight the fact that the optimizations we propose can be extended to simulate the dynamic correlation function of any classical spin Hamiltonian, regardless of the geometry of the lattice and boundary condition. And the idea of fusing memory-bounded calculation into computing-bounded computation and performing on-the-fly computation can be integrated into other scientific computing applications. 

\section*{Acknowledgments}

This work is supported by the U.S. Department of Energy, Office of Science, Basic Energy Sciences under Award No. DE-SC0022216. Part of this work was also supported under Contract DE-AC02-76SF00515, both for the Materials Sciences and Engineering Division and for the Linac Coherent Light Source (LCLS). The numerical calculations were performed using the resources of the National Energy Research Scientific Computing Center (NERSC), a DOE Office of Science User Facility operated under Contract No. DE-AC02-05CH11231.







\end{document}